\documentclass[12pt]{iopart}
\usepackage{graphicx}

\usepackage{iopams} 
\usepackage{txfonts}
\begin{document}

\title[On the failure of Schr\"odinger dynamics in the free expansion of relativistic particles]{Quantifying the failure of Schr\"odinger dynamics in the free expansion of relativistic particles}

\author{Ricardo Ximenes and Fernando Parisio}

\address{Departamento de F\'{\i}sica, Universidade Federal de Pernambuco, 50670-901, Recife, Pernambuco, Brazil}
\ead{parisio@df.ufpe.br}
\vspace{10pt}
\begin{indented}
\item[]February 2016
\end{indented}

\begin{abstract}
It is known that Schr\"odinger equation fails in describing the dynamics of highly energetic particles. We propose to quantify this lack of Lorentz covariance by evaluating the probability for a particle to be measured outside the set of light cones which are compatible to its initial wave function. We consider a simple case of a particle released from a box, which, in turn, is inside a larger container. It is shown that besides the increasing error at relativistic energies, there may be a complete breakdown, with Schr\"odinger dynamics implying in deterministic, superluminal signaling for Lorentz factors above 129. In addition, we give an exact asymptotic expression for the violation in local causality by employing the stationary exponent method, from which the Compton wave length of the particle naturally arises as the relevant scale for the stationary points.
\end{abstract}


%\maketitle

\section{\label{sec:level1}Introduction}
Generally speaking, we know that if the energy of a massive particle is sufficiently high, its predicted nonrelativistic quantum dynamics, as given by Schr\"odinger equation, leads to inconsistent results. But to what extent? Can we quantify, or at least establish a lower bound for the lack of covariance of the Schr\"odinger probability density? Of course, if one is capable of solving the appropriate relativistic equation, this can be done by direct comparison of measurable quantities in the two scenarios. However, this is often not the case. 

Let us start by considering a somewhat deceptive example which, however, illustrates the kind of situation we intend to address. It involves the spreading of free wave packets and goes as follows: Given that the particle's wave function is a Gaussian with initial width $\sigma_0$ at $t=0$, for later times the width will be $\sigma(t)=\sqrt{\sigma_0^2+(\hbar t/2m\sigma_0)^2}$. If the particle's wave length is sufficiently small, $\sigma_0<\hbar/2mc  $, then we have $\sigma(t)>ct$, where $c$ is the speed of light in vacuum. 
Thus, we would have a finite probability to measure the particle in a position that would violate the postulates of special relativity. Although there is nothing in nonrelativistic quantum mechanics that prevents this sort of unphysical prediction, the reasoning is misleading. Although the central region with characteristic scale $\sigma_0$ presents the highest probability density, from the very beginning, the probability to find the particle at any region of space is nonvanishing, for the Gaussian is nonzero everywhere. So, even if we measure the particle outside the light cone associated to the wave-packet {\it centroid}, one can always argue that there was a finite probability for the particle to ``be'' near the detection position all the time.

It is clear from this example that a situation in which one can consistently evaluate to what extent special relativity is being violated by Schr\"odinger dynamics should involve an initially confined system, so that the probability to measure the particle outside the confining region is rigorously null for $t=0$, the confinement conditions posteriorly changing. Confined quantum particles subjected to constraints that become time dependent have been studied in a variety of scenarios. Perhaps the most widely known problem in this class is the Fermi-Ulam bouncer \cite{fermi,ulam,jain} that concerns a particle that moves between two reflective walls, one of which is oscillating. Here we are interested in models with one receding wall, as the ones addressed in \cite{makowski} in the one-dimensional case. In three dimensions there are solutions for expanding cylindrical \cite{cylinder} and spherical \cite{sphere} wells. Of course, if the moving wall has a velocity wich is smaller than the speed of light, then the boundary conditions themselves prevent Schr\"odinger dynamics from predicting violations in local causality. But then, a wall moving at a superluminal velocity is a crying unphysical system. Therefore, for our purposes, the most natural scenarios to address are those in which one of the walls is suddenly removed \cite{aslangul}, as in the free expansion of a classical gas. We will consider these situations in detail in Sec. 3.
\section {Set of compatible light cones}
\begin{figure}
\includegraphics[width=5cm]{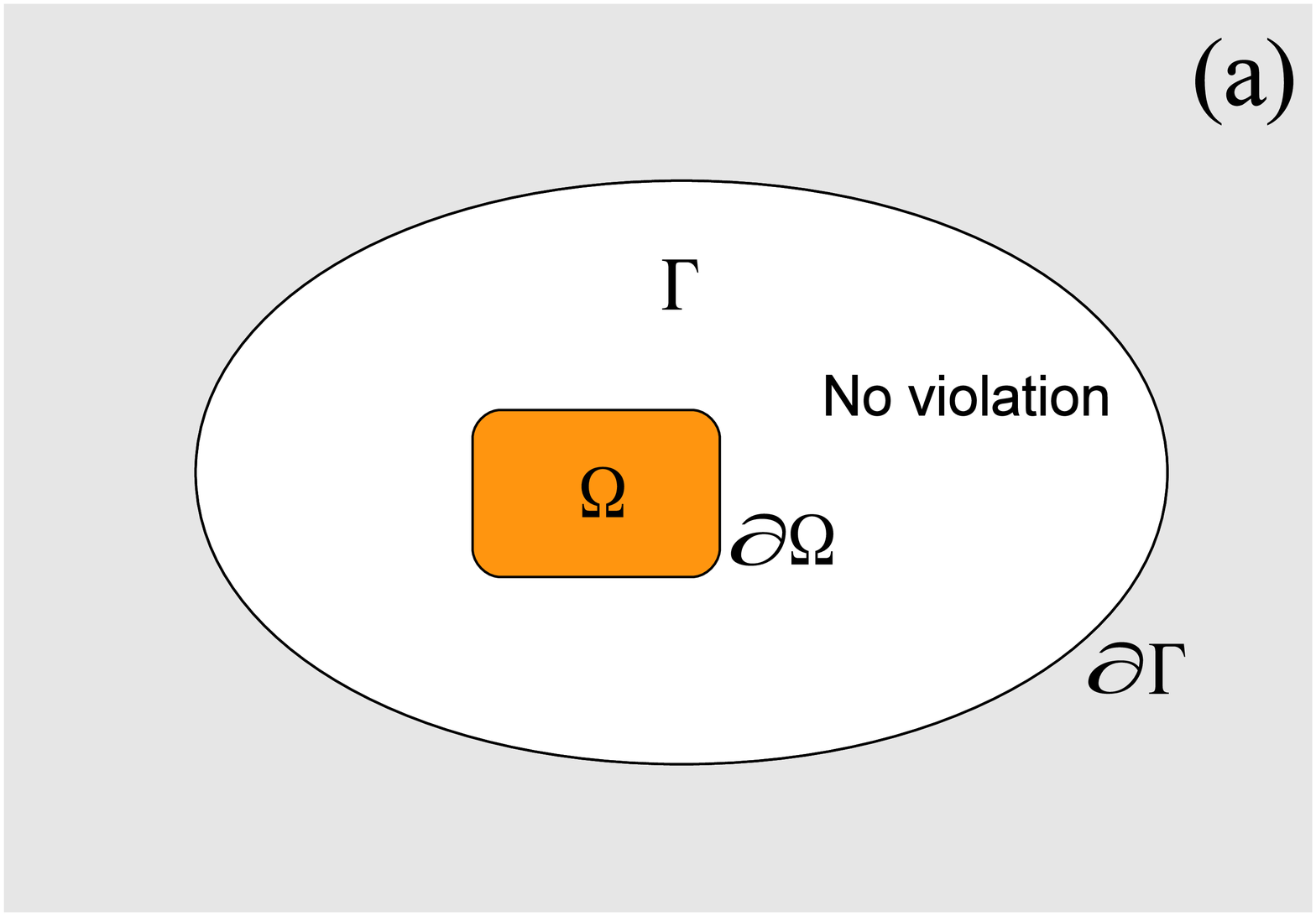}
\includegraphics[width=5cm]{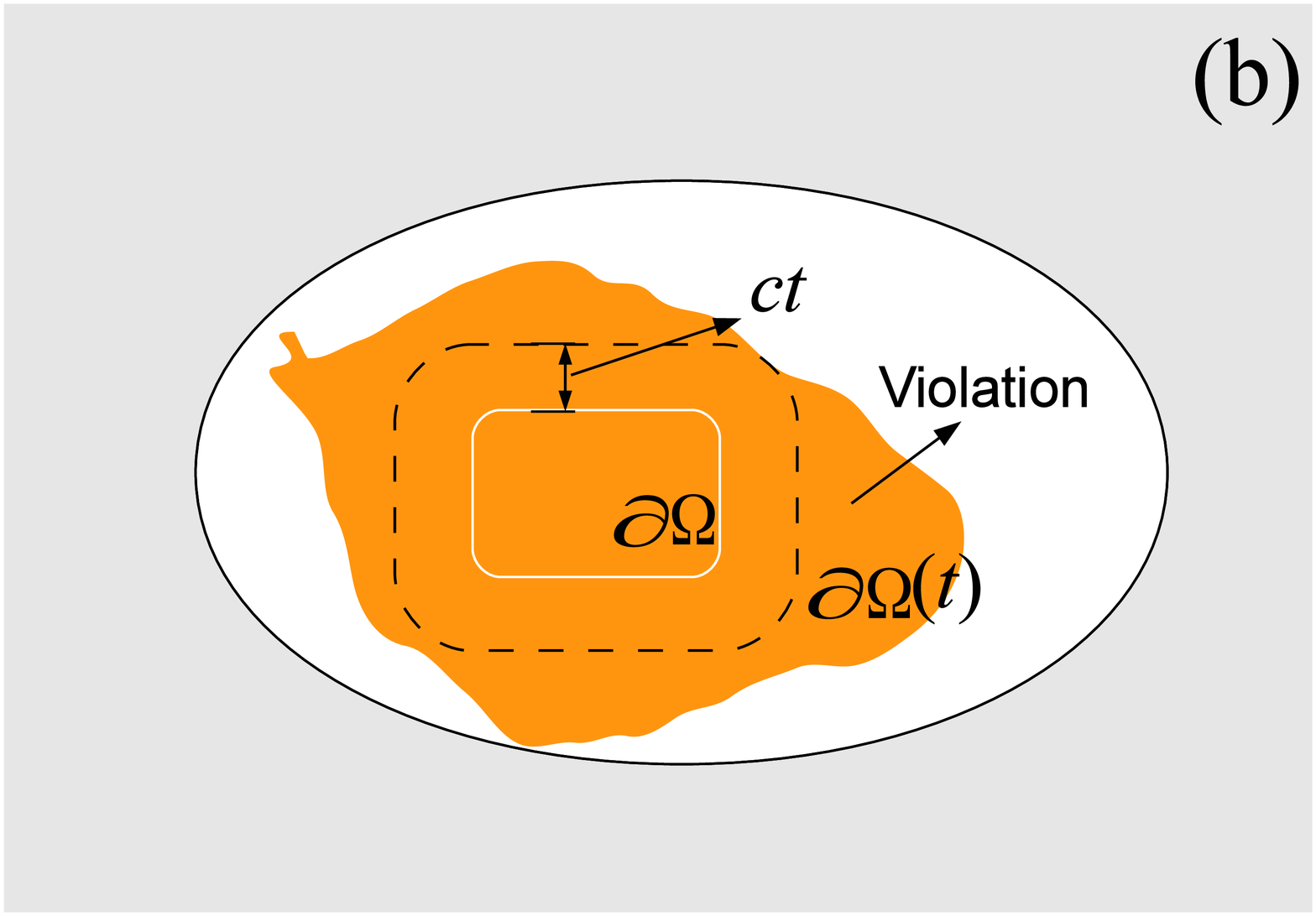}
\includegraphics[width=5cm]{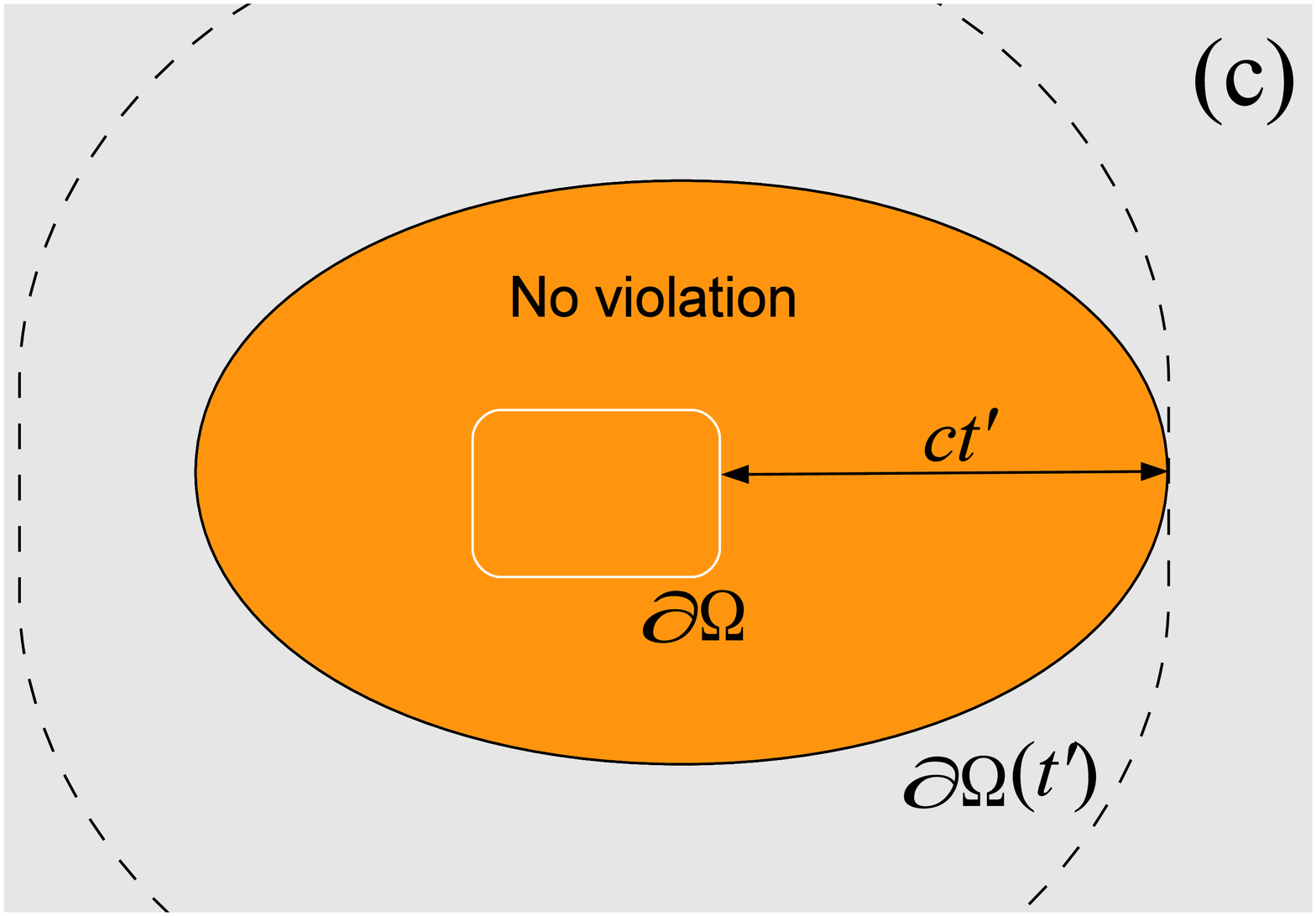}
\caption{Pictorial view of a generic free expansion of a particle's wave function. For times in the interval $(0,t')$ part of the wave function is outside all possible time-like positions the particle could be measured on. }
\label{fig1}
\end{figure}

Let us define what we mean by a set of compatible light cones and then propose our quantifier as a probability of violation in local causality. Let $\Omega$ be a bounded region in $\mathbb{R}^3$ (the reasoning is the same for regions in $\mathbb{R}^2$ and $\mathbb{R}$) and $\partial \Omega$ its outer boundary. Assume that a quantum particle of mass $m$ is confined inside $\Omega$, $\partial \Omega$ corresponding to the walls of a rigid box. For simplicity, let us assume that the particle is in one of the stationary states of this infinite well, say, the fundamental state $\phi_0(\bf r)$. Now, let $\Gamma (\supset \Omega$) be a larger region in space bounded by $\partial \Gamma$ [see Fig. \ref{fig1}(a)]. 

Next, consider that the wall $\partial \Omega$ is suddenly removed (at $t=0$). The wave function $\phi_0(\bf r)$ is no longer an eigenstate of the Hamiltonian and, thus, will evolve according to its expansion in terms of the new eigenfunctions associated with the well delimited by $\partial \Gamma$, $\psi_n(\bf r)$. Therefore, we get $\phi({\bf r},t)=\sum a_n \exp(-i E_n/\hbar) \psi_n(\bf r)$, where $n$ may denote a set of quantum numbers depending on the dimensionality of the system. This situation, as we will see, leads to non-covariant results, since, in general, the previous summation includes modes of {\it arbitrarily high momenta}.

We define $\Omega (t)$ as the points interior to the surface $\partial \Omega(t)$, which is built by the propagation of each point in $\partial \Omega$ in the local normal direction, with the speed of light, see Fig. \ref{fig1}(b). This definition is sufficient for smooth wells (its generalization for wells containing corners is straightforward). Note that, differently from $\partial \Omega$, $\partial \Omega (t)$ does not represent an actual physical constraint for $t\ne 0$, but only a reference surface.
Now one is in position to define the set of compatible light cones as the union of all future light cones referring to the space-time events $\{({\bf r},ct) \,| \,{\bf r} \in \Omega, \;t=0 \}$ which are interior to $\partial \Gamma$.

We finally define the failure of the Schr\"odinger wave function in delivering results compatible with relativistic causality at a time $t$ through the probability ${\cal P}$ to measure the particle outside the set of compatible light cones. That is
\begin{equation}
\label{PV}
{\cal P}(t)=\int_{\Gamma-\Omega (t)}\rho(x,t) d{\bf r}\;,
\end{equation}
where $\rho(x,t)=|\phi({\bf r},t)|^2$ and $d{\bf r}$ is an element of volume. Note that for $t=0$, ${\cal P}=0$ because $\phi({\bf r},0)$ vanishes in $\Gamma- \Omega$ (the set of all points the are in $\Gamma$ and not in $\Omega$). Also, since $\Gamma$ is bounded, there exists a finite time, say $t'$, for which $\Gamma- \Omega(t')=\emptyset$, leading to ${\cal P}=0$ for $t\ge t'$ [Fig. \ref{fig1}(c)].

One important point, that must be clearly understood from the outset, is that there is no relation between genuine Bell nonlocality (which requires entanglement) and the quantity in Eq. (\ref{PV}).
While the former is a true physical effect, the latter is related to the error of Schr\"odinger dynamics in relativistic situations due to its Galilean covariance. In other words, the probability ${\cal P}$ gives a lower bound for the amount of {\it fake} nonlocality that arises in non-relativistic quantum mechanics. It is a lower bound because the part of the wave packet which is inside the set of compatible light cones not necessarily yields precise predictions on experimental results. However, for this part of the wave function at least local causality is preserved.
\section{A simple example}
In this work we will address what could be named ``the free expansion of a single particle in one dimension'' \cite{aslangul}. It consists of the simplest instance of the general process described in the previous section. We consider a particle in an infinite one-dimensional well of width $a$ which, at $t=0$, is released to expand within a larger well with size $\Lambda a$, $\Lambda>1$. Therefore, in this case, $\Omega$ and $\Gamma$ correspond to the intervals $(0,a)$ and $(0,\Lambda a)$, respectively. The boundary of $\Omega$ is given by the two points $\{0,a\}$ and $\partial \Omega (t)=\{0,a+ct\}$. 
In addition,
\begin{equation}
\label{phi0}
\phi_0(x)=\sqrt{\frac{2}{a}} \sin\left(\frac{\pi x}{a}\right)\;, 
\end{equation}
is the stationary solution for $t\le 0$, while the basis in terms of which $\phi_0$ has to be decomposed for $t>0$ is
\begin{equation}
\psi_n(x)=\sqrt{\frac{2}{\Lambda a}} \sin\left(\frac{n \pi x}{\Lambda a}\right)\;.
\end{equation}
The exact solution for the Schr\"odinger dynamics of this problem appeared in this journal in 2008 \cite{aslangul}. The wave function for an arbitrary time $t$ reads
\begin{equation}
\label{exac}
\phi(x,t)=\frac{i\Lambda}{\pi}\sqrt{\frac{2}{a}}\sum_n \frac{1}{n^2-\Lambda^2}\sin\left( \frac{n\pi}{\Lambda}\right) \exp\left( \frac{i n \pi x}{\Lambda a} -\frac{i2 \pi n^2t}{ T_{\Lambda}} \right)
\end{equation}
where the sum runs from $-\infty$ to $+\infty$ and
\begin{equation}
T_{\Lambda}=\frac{4m\Lambda^2 a^2}{\pi \hbar}\;,
\end{equation}
is the revival time in the larger wall. As we will se, this is a relevant time scale in our analysis, especially due to the symmetry property \cite{styer}:
\begin{equation}
\label{sym}
\phi(x,T_{\Lambda}/2)=-\phi(\Lambda a-x,0)\;.
\end{equation}
\subsection{The relativistic regime}
\begin{figure}
\includegraphics[width=5cm]{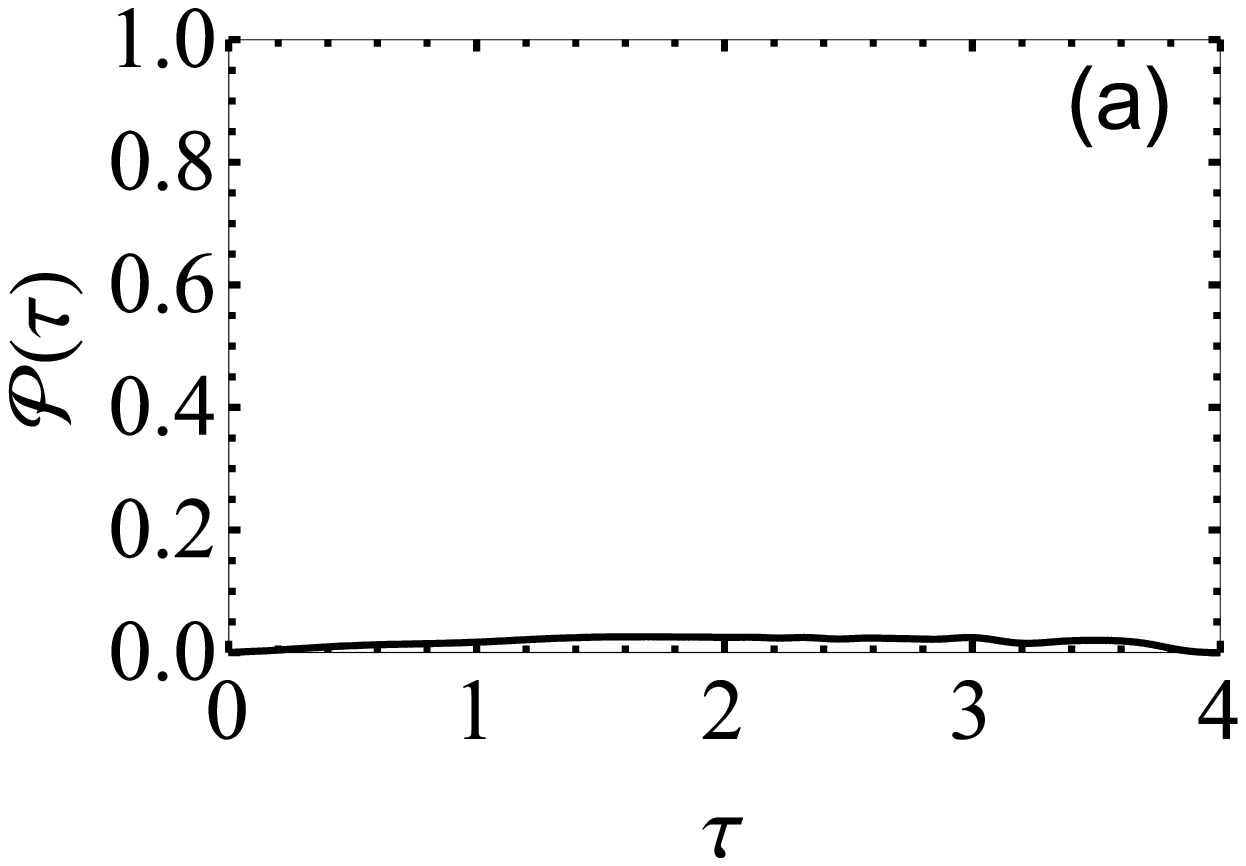}
\includegraphics[width=5cm]{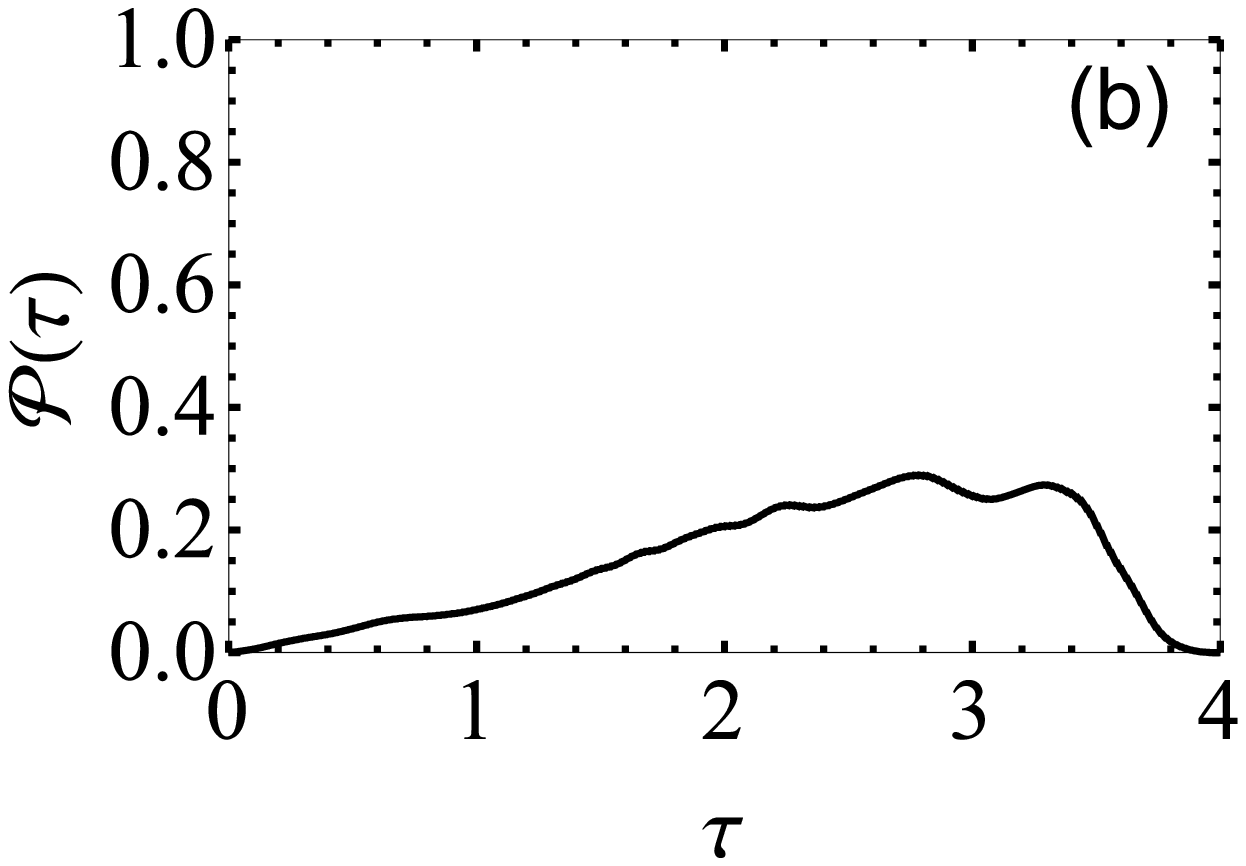}
\includegraphics[width=5cm]{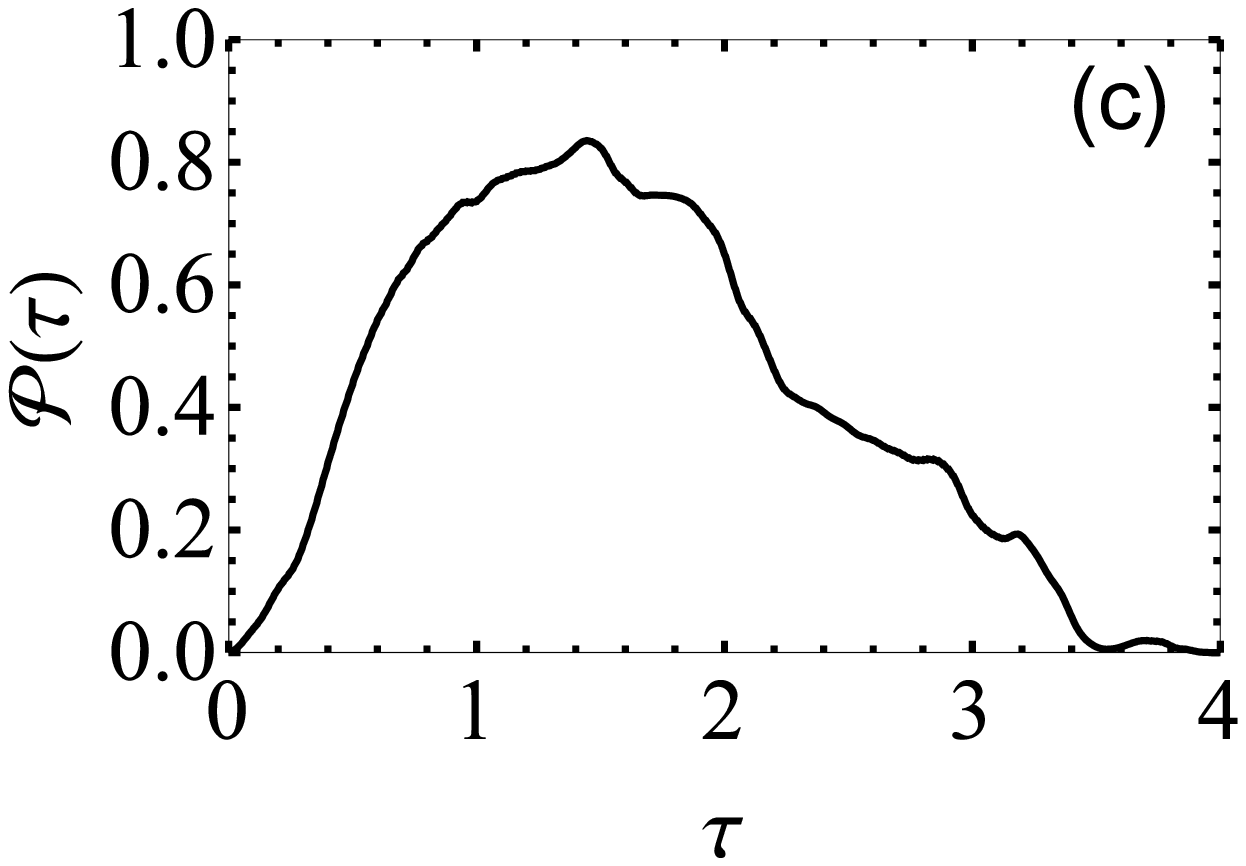}\\
\includegraphics[width=5cm]{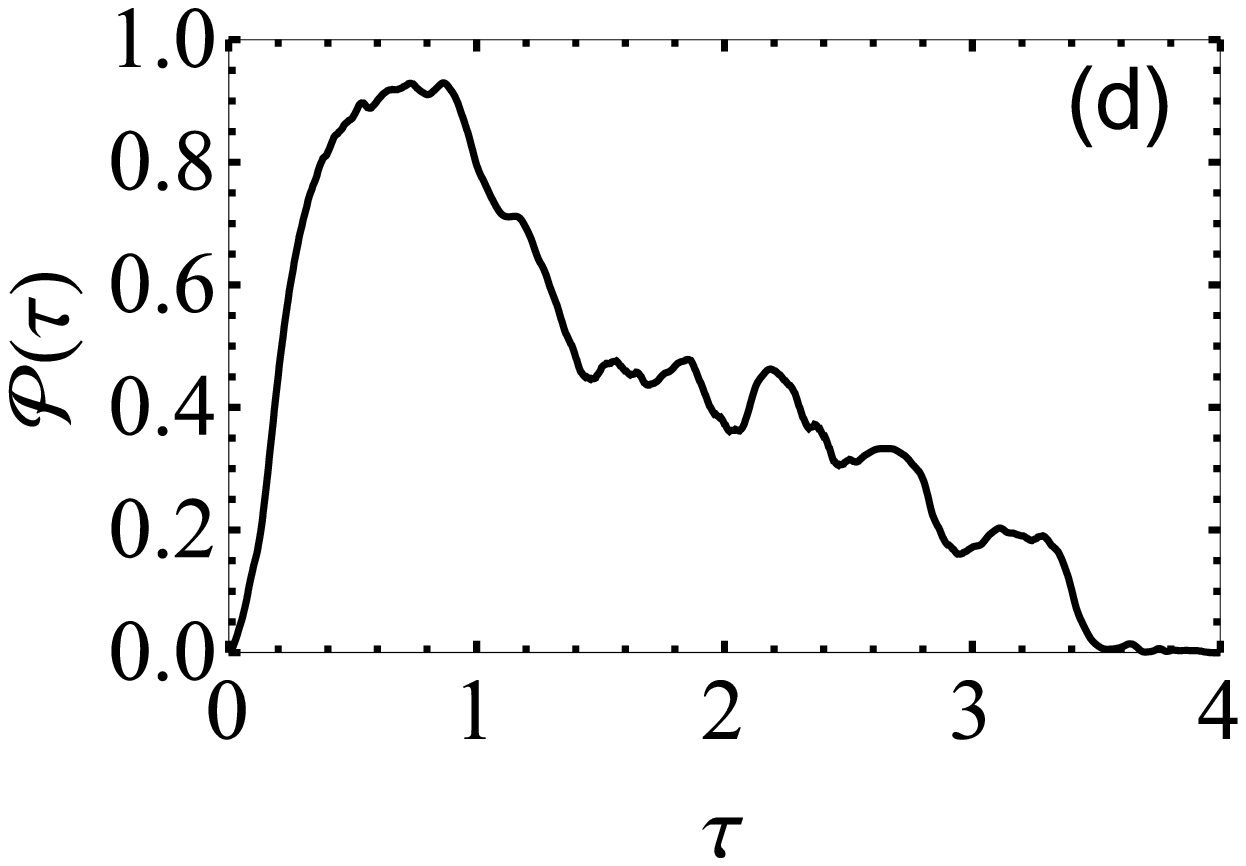}
\includegraphics[width=5cm]{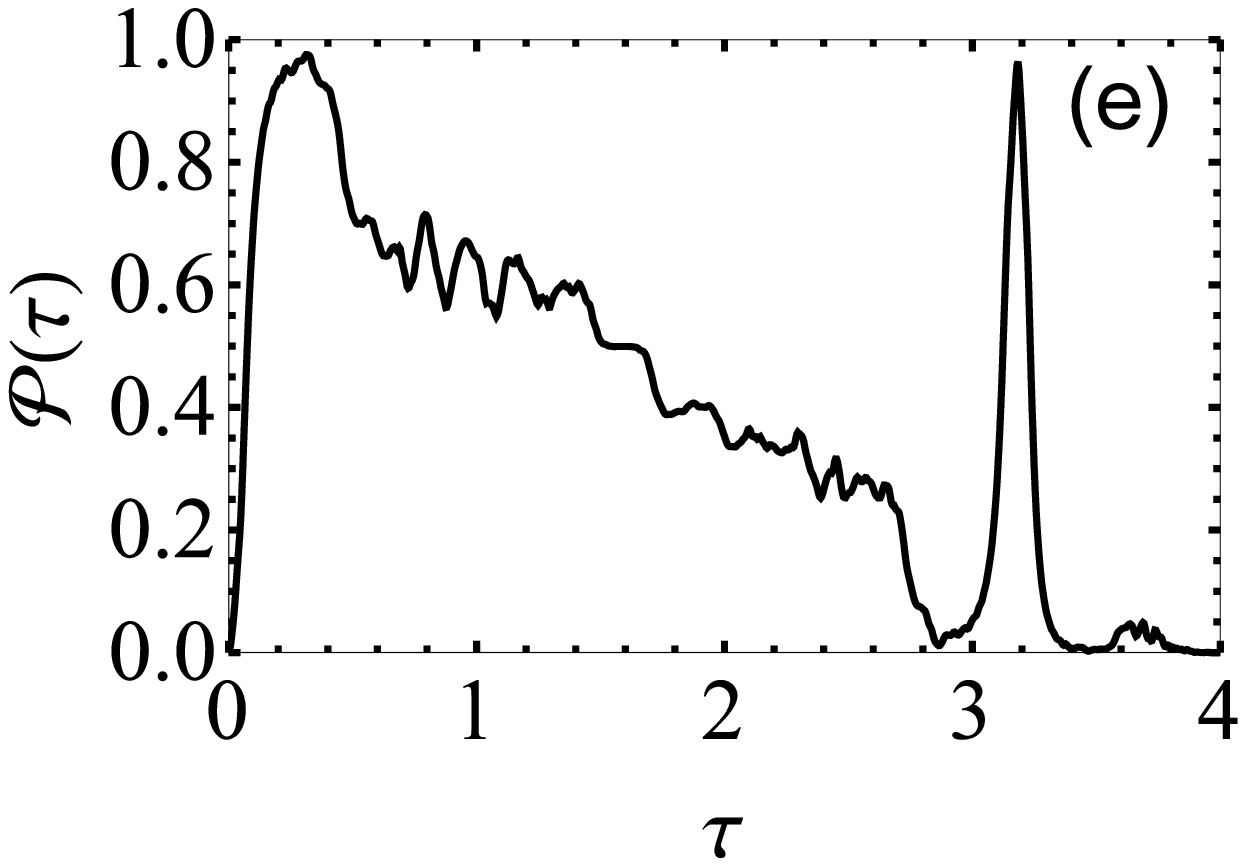}
\includegraphics[width=5cm]{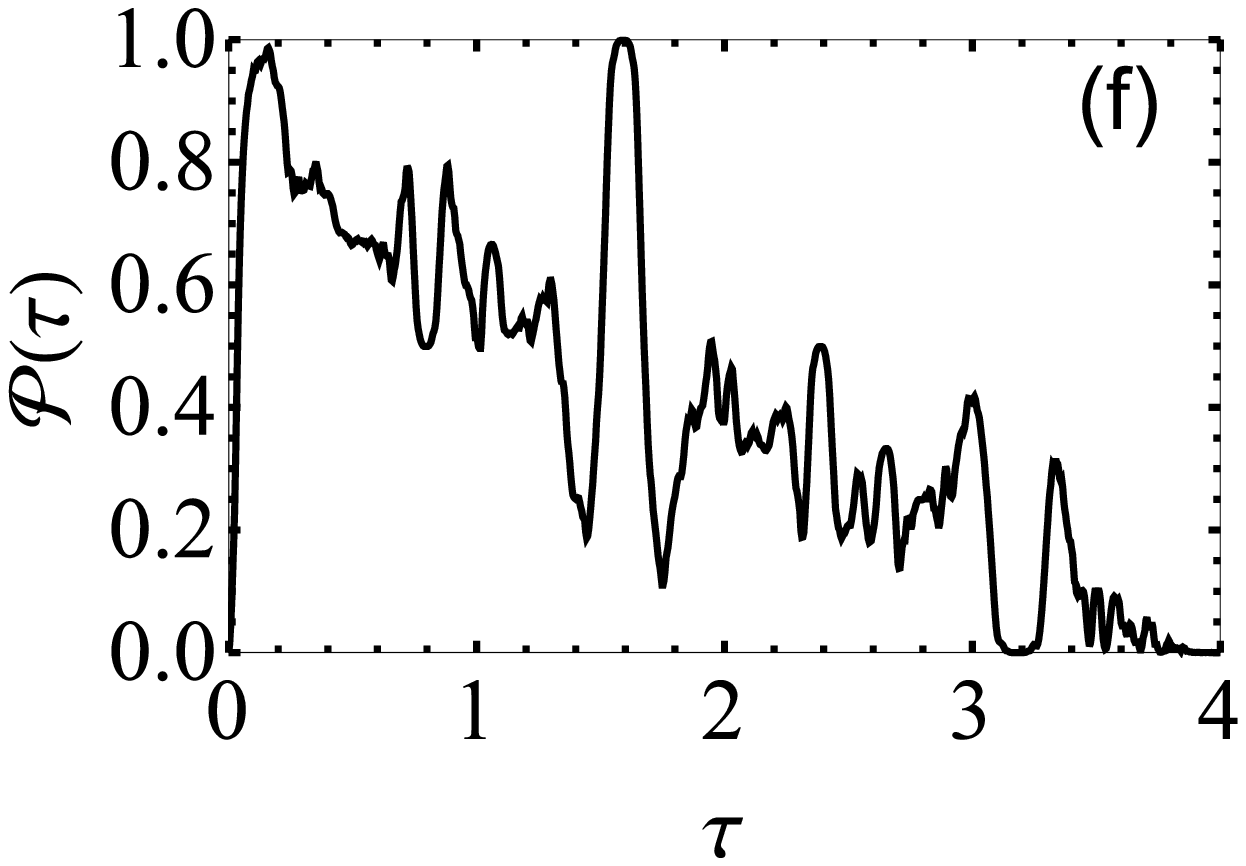}
\caption{Probability of violation as a function of the dimensionless time $\tau=ct/a$ for different box sizes. The expansion factor is the same in all panels ($\Lambda=5$). The probability of violation is less than $0.03$ for $a=4\lambdabar_c$ (a). Failure of Schr\"odinger dynamics becomes increasingly likely for decreasing values of $a$: $a=2.5\lambdabar_c$ (b), $a=\lambdabar_c$ (c), $a=0.5 \lambdabar_c$ (d), $a=0.2\lambdabar_c$ (e), and $a=0.1\lambdabar_c$ (f). In the last two plots there is a second, sharp peak which attains the maximal value of $1$ in (f).}
\label{fig2}
\end{figure}

\begin{figure}
\includegraphics[width=5cm]{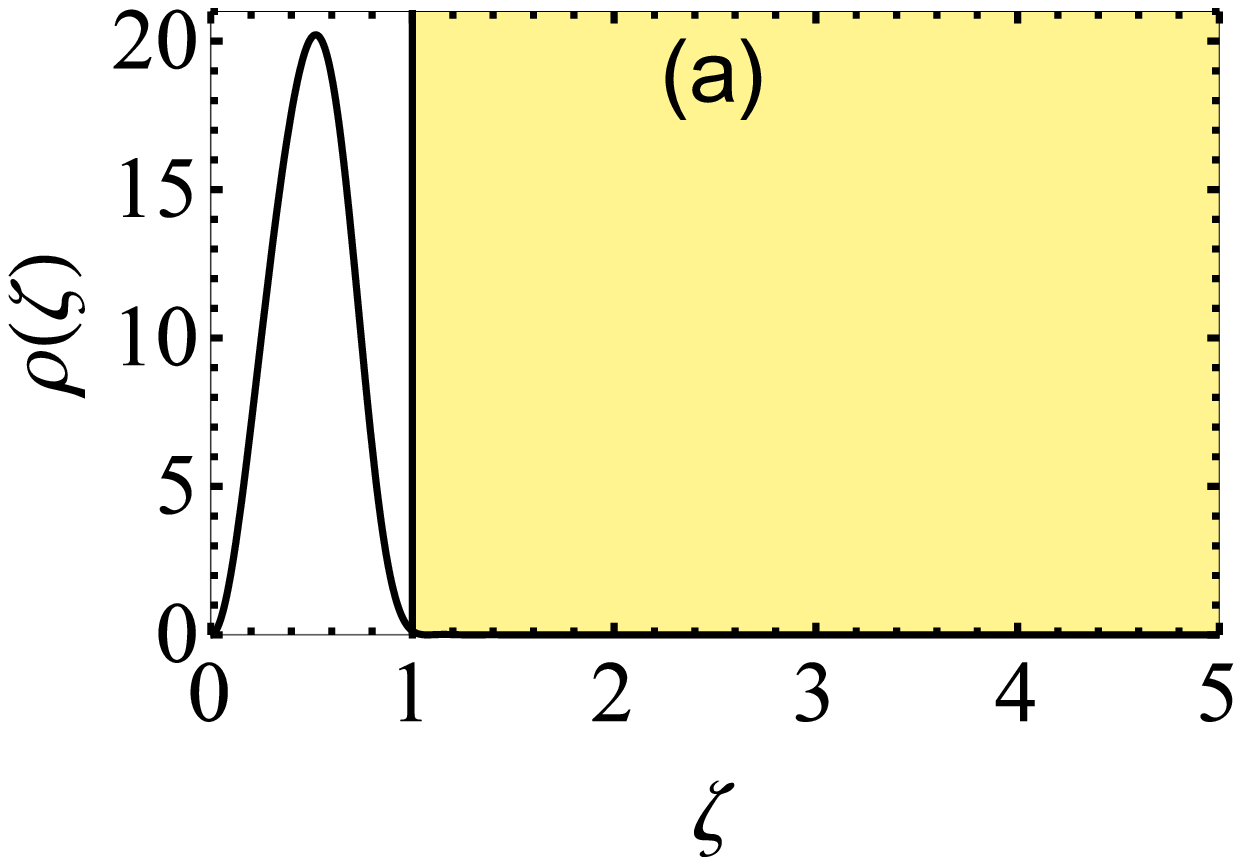}
\includegraphics[width=5cm]{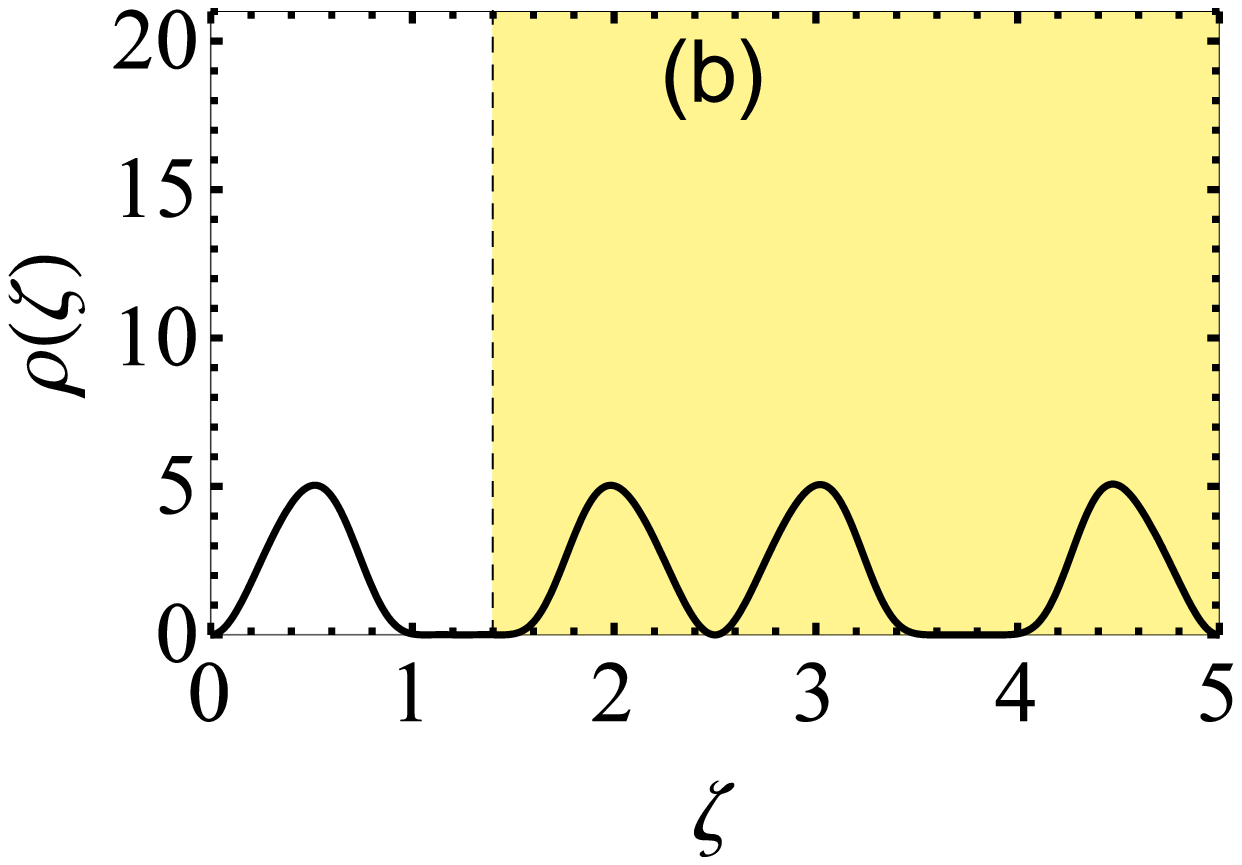}
\includegraphics[width=5cm]{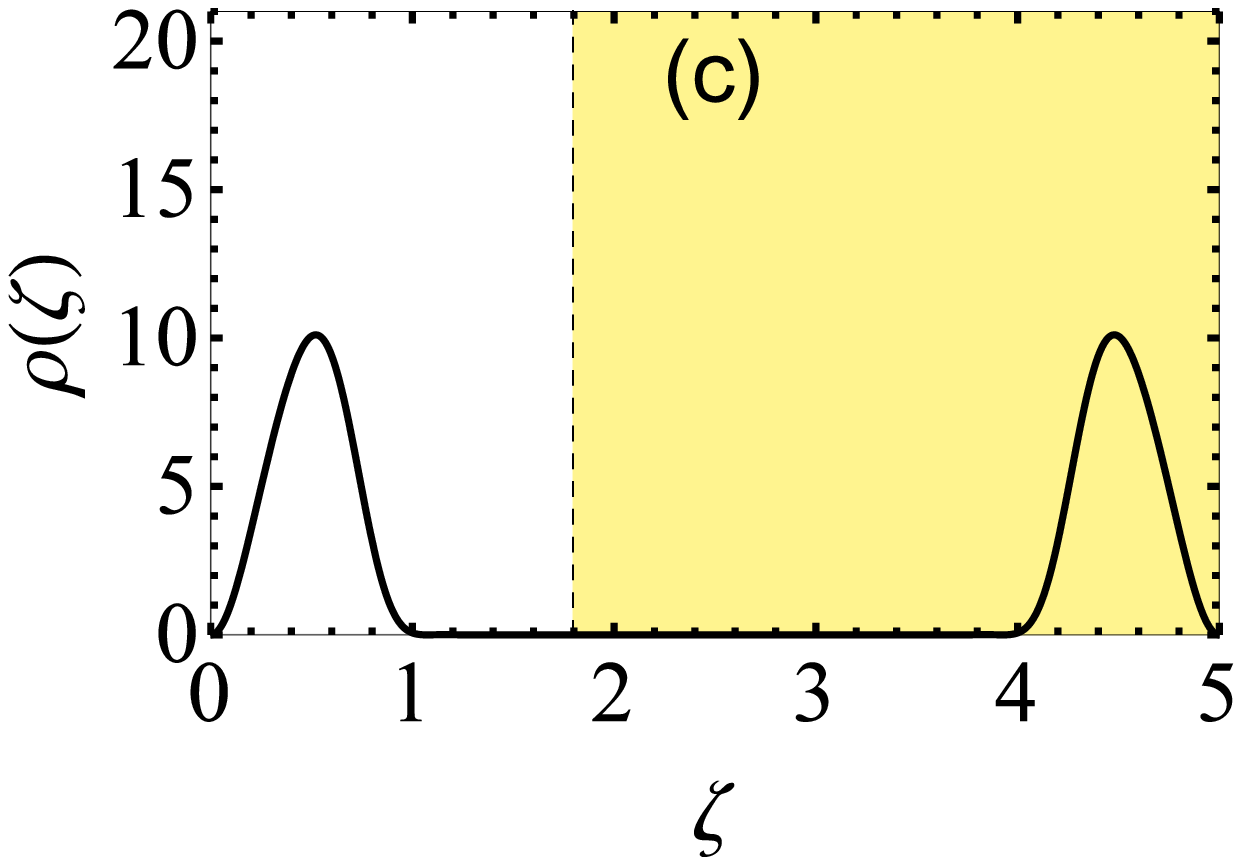}\\
\includegraphics[width=5cm]{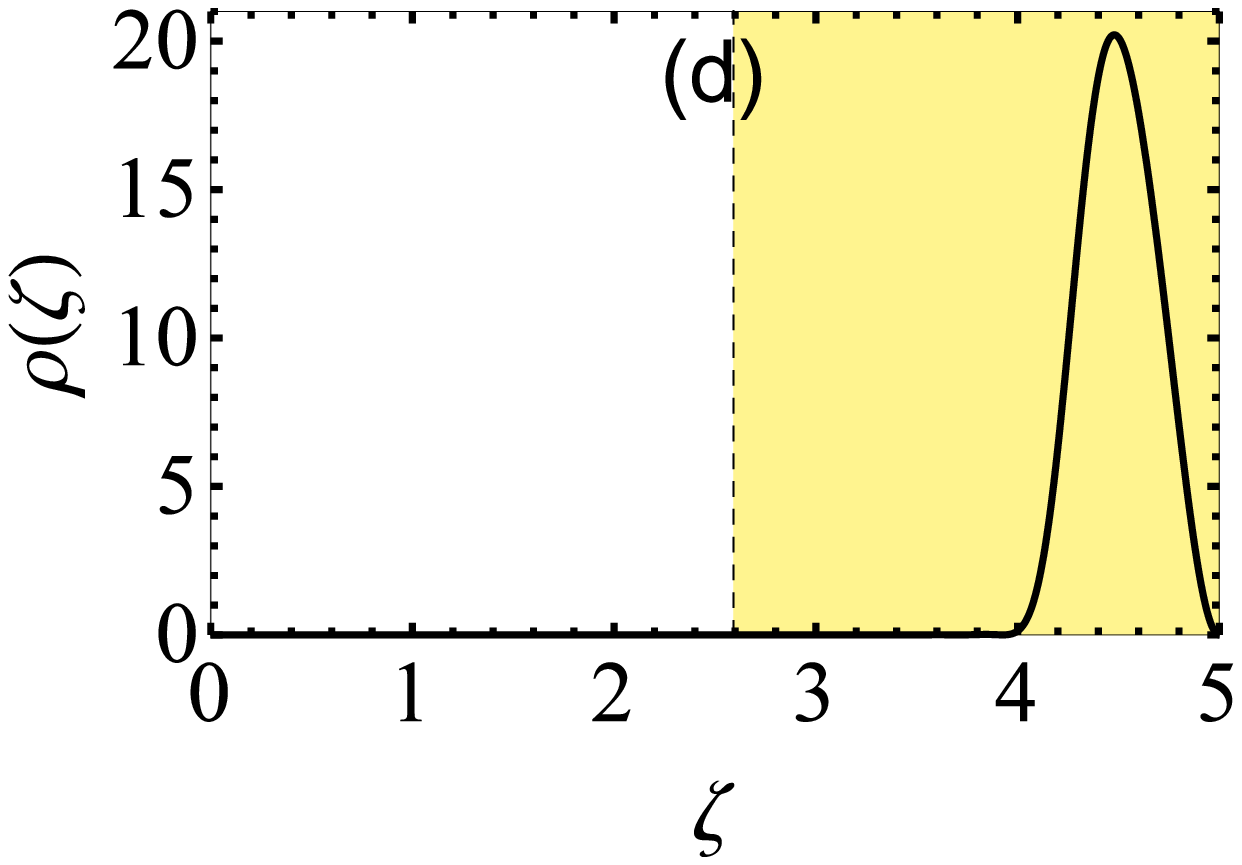}
\includegraphics[width=5cm]{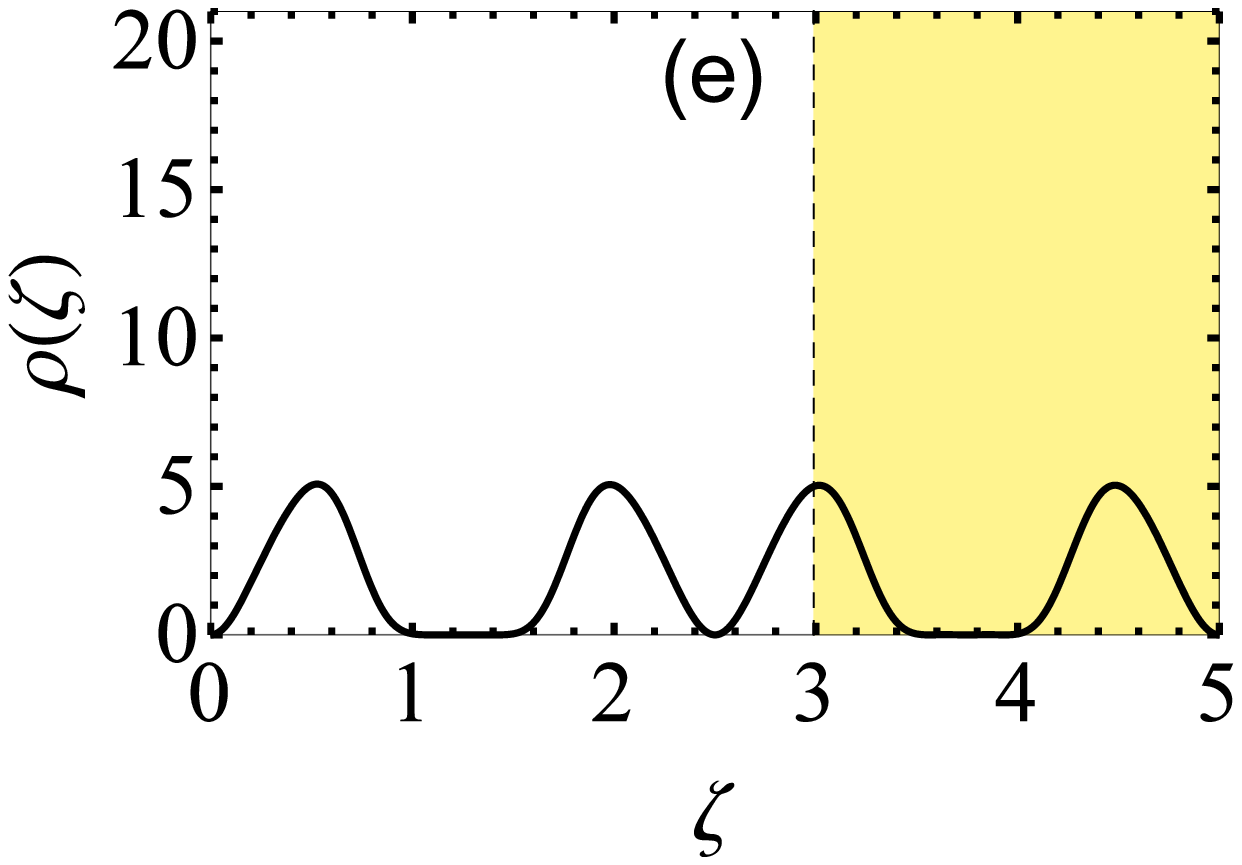}
\includegraphics[width=5cm]{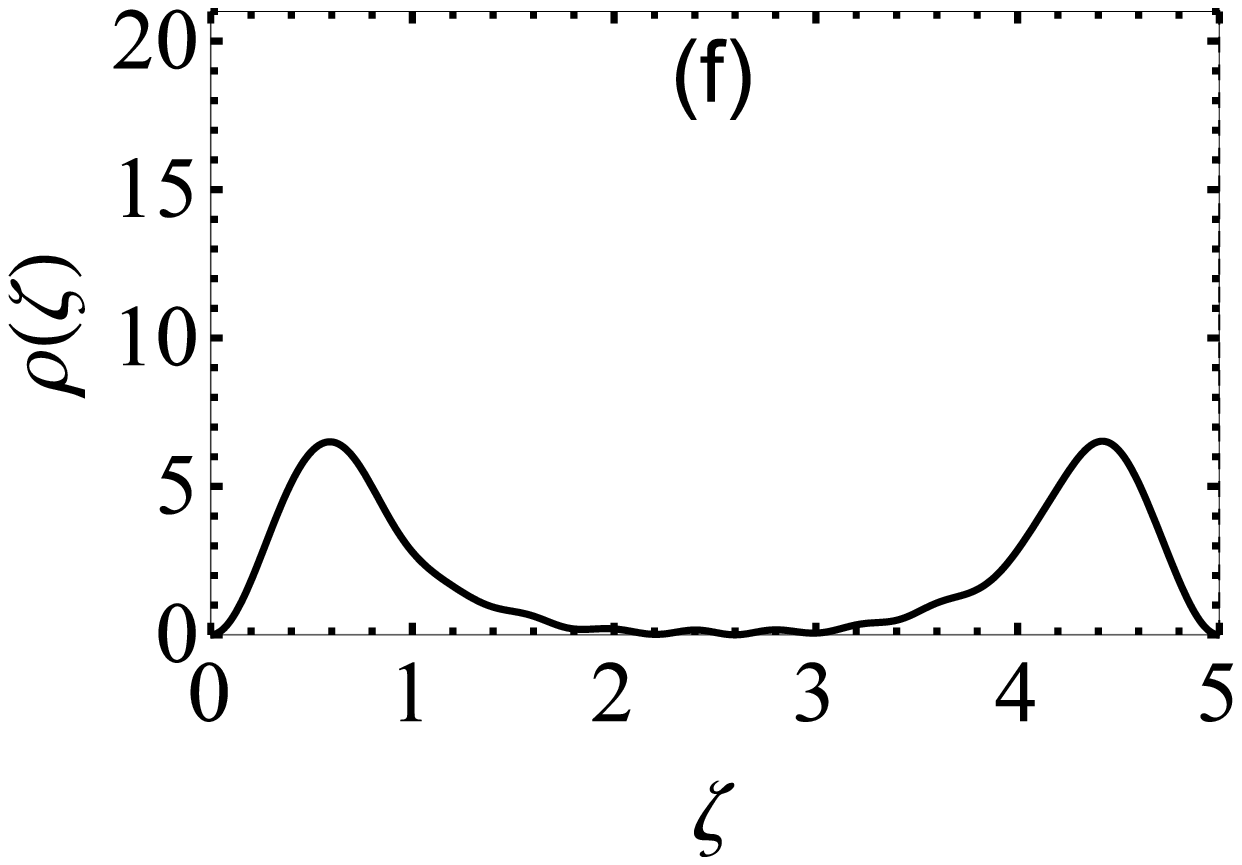}
\caption{Probability density $\rho$ versus $\zeta=x/a$ for selected times. In all panels $a=0.1 \lambdabar_c$ and $\Lambda=5$. The times are $t=0$ (a), $t=T_{\Lambda}/8$ (b), $t=T_{\Lambda}/4$ (c), $t=T_{\Lambda}/2$ (d), $t=5T_{\Lambda}/8$ (e), and $t=(\Lambda a-a)/c$ (f).}
\label{fig3}
\end{figure}

We are interested in situations where the particle's kinetic energy $E_0=\pi^2\hbar^2/2ma^2$ (the energy of the fundamental state of the inner box) is relevant in comparison to its rest energy ${\cal E}_0=mc^2$. The relation between the initial size of the well and the particle's Lorentz factor can be conveniently expressed as
\begin{equation}
\label{lorentz}
\frac{1}{\sqrt{1-v^2/c^2}}\equiv \gamma=1+\frac{\pi^2}{2}\left( \frac{a}{\lambdabar_c}\right)^{-2}\;,
\end{equation}
where $\lambdabar_c=\hbar/mc$ is the particle's Compton wave length and $v$ is the velocity of a classical particle with kinetic energy $E_0$ [$= (\gamma-1)mc^2$]. In figure \ref{fig2} we show numerically calculated ${\cal P}$, Eq. (\ref{PV}), as a function of the dimensionless time $\tau=ct/a$ for different system sizes. The expansion factor was kept constant, $\Lambda=5$, so that what is being varied from panel to panel is a global scale factor, and, thus, the absolute size of the whole system. Also ${\cal P}(0)={\cal P}(\tau)=0$, $\tau \ge \tau' = 4$. For $a=4\lambdabar_c$ ($\gamma=1+\pi^2/32\approx 1.31$, $v\approx 0.64c $) the failure of Galilean covariance to preserve local causality in quantum mechanics is very small, as can be seen in figure \ref{fig2} (a). Even for such a fast particle the chances that non-relativistic quantum mechanics would predict a violation in local causality is no larger than $3\%$. As the size of the initial confinement region is decreased, ${\cal P}$ starts to display a broad peak which is related to the initial, faster-than-light ballistic motion of the wave packet soon after the change in the boundary conditions. After that ${\cal P}$ (unsteadily) decreases to zero at $\tau =4$ for $a=2.5 \lambdabar_c$ (b), $a= \lambdabar_c$ (c), and $a=0.5 \lambdabar_c$ (d). In the last two panels, $a=0.2 \lambdabar_c$ (e) and $a=0.1 \lambdabar_c$ (f), a second, prominent peak shows up.
This constitutes an extreme symptom that Schr\"odinger dynamics presents when even more energetic particles are considered. It consists in the fact that, according to Eq. (\ref{sym}), at $t=T_{\Lambda}/2$ the wave-packet shape is exactly restored as it was as at $t=0$. This time, however, the probability density is non-vanishing only between $\Lambda a-a<x<\Lambda a$. So, if at this moment the light front associated to the removal of the constraint is behind $x=\Lambda a-a$, then the probability of violation becomes exactly $1$ [${\cal P}(t=T_{\Lambda}/2)=1$], and, thus, we have a $100\%$ failure of the Schr\"odinger dynamics to keep up local causality. Note that if such a phenomenon could be indeed produced, one would have the ability to {\it deterministically} send superluminal information from region $[0,a]$ to region $[\Lambda a -a, \Lambda a]$. 
This situation is depicted in the snapshots of figure \ref{fig3} for the same set of parameters used in figure \ref{fig2} (f). The probability densities are shown as functions of the dimensionless distance $\zeta=x/a$. At $t=0$, Fig. \ref{fig3} (a), the wave function is given by equation (\ref{phi0}) for $0<\zeta<1$, vanishing elsewhere. The other snapshots correspond to $t=T_{\Lambda}/8$ (b), $t=T_{\Lambda}/4$ (c), $t=T_{\Lambda}/2$ (d), $t=5T_{\Lambda}/8$ (e), and $t=(\Lambda a-a)/c$ (f). Note that for $t=T_{\Lambda}/2$ the wave packet is {\it totally} contained in a space-like region, leading to the second peak, ${\cal P}=1$, in figure \ref{fig2} (f).

This radical breakdown would occur whenever the time required for light to propagate between $x=a$ and $x=\Lambda a -a$ is larger than the ``specular-revival'' time, that is, 
\begin{equation}
\frac{\Lambda a-2a}{c} \ge \frac{T_{\Lambda}}{2}\;\; \mbox{or}\;\; \frac{2a}{\pi \lambdabar_c}\Lambda^2-\Lambda+2 \le 0\;.
\end{equation}
Therefore, a necessary condition on the expansion parameter $\Lambda$ would be
\begin{equation}
\label{interval}
\frac{\pi \lambdabar_c}{4 a}\left( 1- \sqrt{1-\frac{16 a}{\pi \lambdabar_c}}\right)\le \Lambda \le \frac{\pi \lambdabar_c}{4 a}\left( 1+ \sqrt{1-\frac{16 a}{\pi \lambdabar_c}}\right)\;,
\end{equation}
which makes evident the existence of a threshold initial confinement bellow which total failure may happen (depending on wether or not the previous condition on $\Lambda$ is satisfied)
\begin{equation}
\label{condition}
a\le \frac{\pi}{16}\lambdabar_c \approx 0.196 \lambdabar_c\;.
\end{equation}
By replacing this critical value in (\ref{lorentz}) we get the minimal Lorentz factor that would lead to this breakdown: $\gamma = 129$. Note that, among the situations addressed in Fig. \ref{fig2}, the only which satisfies conditions (\ref{condition}) and (\ref{interval}), that reads $2.36 \le \Lambda \le 13.36$, is that of panel (f), since $a/\lambdabar_c=0.1<0.196$ and $\Lambda=5$. Note, in addition, that Fig. \ref{fig2} (e) refer to $a/\lambdabar_c=0.2$ which is just above the threshold in (\ref{condition}). Thus, although we still see a sharp peak, it does not attain the maximum value $1$.
\section{Release in free space: exact asymptotic results}
\begin{figure}
\begin{center}
\includegraphics[width=6cm]{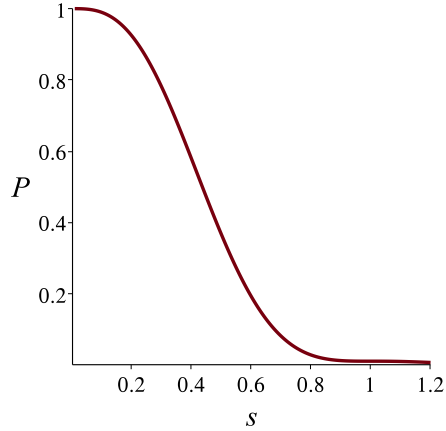}
\end{center}
\caption{Asymptotic probability $P$ as a function of the dimensionless confinement size $s=a/\lambdabar_c$. For $s=1$ we have about $1\%$ of asymptotic violation in local causality.}
\label{fig4}
\end{figure}

A natural question that arises is what happens if there is no outer box whatsoever, the situation corresponding to the limit $\Lambda\rightarrow \infty$.
We proceed to show that, in this case, the probability of violation can be determined analytically for sufficiently large $t$. It turns out that $\lim_{t\rightarrow \infty}{\cal P}(t)$ depends only on the ratio of $a$ and the Compton wave-length of the released particle.
In this case, the exact propagator is known \cite{aslangul} and the wave function reads:
\begin{equation}
\phi(x,t)=\frac{i \sqrt{2}}{a^{3/2}}\int_{-\infty}^{+\infty}dk\frac{\sin(ka)}{k^2-(\pi/a)^2}\;e^{ikx}e^{-i\hbar k^2t/2m}\;.
\end{equation}
Thus, the probability of violation is 
\begin{equation}
{\cal P}(t)=1-\frac{2ct}{a^3}\int_0^{1+a/ct}dy \chi^{*}(y,t)\chi(y,t)
\end{equation}
where $x=yct$ and
\begin{equation}
\chi(y,t)=\int_{-\infty}^{+\infty} dk \left[\frac{\sin(ka)}{k^2-(\pi/a)^2}\right]\;e^{it F }\;,
\label{chi}
\end{equation}
where $F=yck-\hbar k^2/2m$. 
In the asymptotic limit $t\rightarrow \infty$, expression (\ref{chi}) becomes the archetypal integral for which the stationary-phase method can be applied. 
Note that there is no singularity in the (slowly-varying) pre-factor between the square brackets in the above integral, since, as $k\rightarrow \pi/a$, it goes to $-a^2/2\pi$. 
The stationary-phase condition, $\partial F/\partial k=0$, gives 
\begin{equation}
k_{stationary}=\frac{mcy}{\hbar}=\frac{y}{\lambdabar_c}\;.
\end{equation}
Note that $\lambdabar_c$ naturally arises as the scale connecting the stationary wave number and the dimensionless variable $y$. The final, lower order result for the asymptotic violation probability is
\begin{equation}
{\cal P}(t\rightarrow\infty)\equiv P\left( s\right)=1-4\pi\int_0^{s} \frac{\sin^2\theta}{(\theta^2-\pi^2)^2} d\theta\;,
\end{equation}
with $s=a/\lambdabar_c$. This function is depicted in figure \ref{fig4}. It is easy to check that for $s\rightarrow \infty$, $P\rightarrow 0$, as it should be. After some manipulations this result can be written in terms of tabulated functions as:
\begin{eqnarray}
\nonumber
P(s)=1-\frac{1}{\pi}\left[ {\rm Si}(4\pi s-2\pi)+{\rm Si}(4\pi s+2\pi)\right]+\frac{4s}{\pi^2}\frac{\sin^2(2\pi s)}{4s^2-1}\\
-\frac{1}{2\pi^2}\left[ {\rm Ci}(4\pi s-2\pi)-\ln(2\pi s-\pi)-{\rm Ci}(4\pi s+2\pi)+\ln(2\pi s+\pi)\right]\;,
\end{eqnarray}
where Ci and Si denote the cosine and sine integral functions, respectively \cite{abramowitz}.
It is worth noticing that, for small $s$ (high energy), we have 
\begin{equation}
P(s)\sim 1-\frac{4}{3} (2s)^3+ O(s^5)\;,
\end{equation}
so that the first non-constant term is of cubic order. The probability starts to considerably decrease at $s\approx 0.2$ as can be seen in Fig. \ref{fig4},
which is consistent with equation (\ref{condition}).
\section{Concluding remarks}
The fact that Schr\"odinger equation is covariant under Galilean transformations leads to erroneous predictions in the relativistic limit. We showed that a possible way to be more quantitative about this point is to determine the probability of a particle to be measured outside the light cones that are compatible to its initial wave function, according to Schr\"odinger equation. 

The ultra-relativistic limit can be sharply defined for the specific problem we addressed, as that encompassing the situations in which the particle's initial confinement satisfies $a\le \pi/ 16 \lambdabar_c$, corresponding to a Lorentz factor of $\gamma=129$. In this regime, depending on the value assumed by the expansion parameter $\Lambda$, Schr\"odinger dynamics would allow for {\it deterministic} superluminal signaling. This this serious physical impossibility arises in the nonrelativistic description due to a competition between the two time scales $T_{\Lambda}/2$ and $(\Lambda a -2a)/c$.

We also derived a closed analytical expression for the asymptotic probability of violation in the case of a particle released to propagate in a semi-infinite space. In this case, the only relevant parameter is the ratio between the size of the confinement region and the Compton wave length of the particle, which naturally emerges in the stationary-phase evaluation.

It would be interesting to cope with the same class of problems in higher dimensional systems, for instance, spherical and cylindrical boxes \cite{cylinder, sphere}. A relevant question is: Given that the particle's energy in the inner box is kept constant, what is the behavior of the probability of violation ${\cal P}$ as the system's dimensionality grows from 1 to 3? 
\ack
R. X. would like to thank Leonardo Cabral for his help with the numerical calculations. Financial support from Conselho Nacional de Desenvolvimento Cient\'{\i}fico e Tecnol\'ogico (CNPq) through its program INCT-IQ, Coordena\c{c}\~ao de Aperfei\c{c}oamento de Pessoal de N\'{\i}vel Superior (CAPES), and Funda\c{c}\~ao de Amparo \`a Ci\^encia e Tecnologia do Estado de Pernambuco (FACEPE) is acknowledged.

\end{document}